\documentclass[12pt,thmsa,onecolumn,ukenglish]{article}
\usepackage{amsmath}
\usepackage{sw20lart}

\input{tcilatex}
\begin{document}

\title{Time-Fractional KdV Equation Describing the Propagation of
Electron-Acoustic Waves in plasma }
\date{}
\author{S. A. El-Wakil, E. M. Abulwafa, \and E. K. El-shewy and A. A.
Mahmoud \\
Theoretical Physics Research Group, Physics Department, \\
Faculty of Science, Mansoura University, Mansoura 35516, Egypt}
\maketitle

\begin{abstract}
The reductive perturbation method has been employed to derive the
Korteweg-de Vries (KdV) equation for small but finite amplitude
electron-acoustic waves. The Lagrangian of the time fractional KdV equation
is used in similar form to the Lagrangian of the regular KdV equation. The
variation of the functional of this Lagrangian leads to the Euler-Lagrange
equation that leads to the time fractional KdV equation. The
Riemann-Liouvulle definition of the fractional derivative is used to
describe the time fractional operator in the fractional KdV equation. The
variational-iteration method given by He is used to solve the derived time
fractional KdV equation. The calculations of the solution with initial
condition $A_{0}\sec $h$(cx)^{2}$ are carried out. The result of the present
investigation may be applicable to some plasma environments, such as the
Earth's magnetotail region .

\begin{description}
\item $\boldsymbol{Keywords:}$ Electron-acoustic waves; Euler-Lagrange
equation, Riemann-Liouvulle fractional derivative, fractional KdV equation,
He's variational-iteration method.

\item \textbf{PACS:} 05.45.Df, 05.30.Pr.
\end{description}
\end{abstract}

\section{Introduction}

\smallskip Because most classical processes observed in the physical world
are nonconservative, it is important to be able to apply the power of
variational methods to such cases. A method used a Lagrangian that leads to
an Euler-Lagrange equation that is, in some sense, equivalent to the desired
equation of motion. Hamilton's equations are derived from the Lagrangian and
are equivalent to the Euler-Lagrange equation. If a Lagrangian is
constructed using noninteger-order derivatives, then the resulting equation
of motion can be nonconservative. It was shown that such fractional
derivatives in the Lagrangian describe nonconservative forces [1, 2].
Further study of the fractional Euler-Lagrange can be found in the work of
Agrawal [3, 4], Baleanu and coworkers [5, 6] and Tarasov and Zaslavsky [7,
8]. During the last decades, Fractional Calculus has been applied to almost
every field of science, engineering and mathematics. Some of the areas where
Fractional Calculus has been applied include viscoelasticity and rheology,
electrical engineering, electrochemistry, biology, biophysics and
bioengineering, signal and image processing, mechanics, mechatronics,
physics, and control theory [9]. 

On the other hand, electron acoustic waves (EAWs) have been observed in the
laboratory when the plasma consisted of two species of electrons with
different temperatures, referred to as hot and cold electrons [10], or in an
electron ion plasma with ions hotter than electrons [11]. Also its
propagation plays an important role not only in laboratory but also in space
plasma. For example, Bursts of broadband electrostatic noise (BEN) emissions
have been observed in auroral and other regions of the magnetosphere, e.g.
polar cusp, plasma sheet boundary layer (PSBL). \ see [12]. There are
different methods to study nonlinear systems [13]. Washimi and Taniti [13]
were the first to use reductive perturbation method to study the propagation
of a slow modulation of a quasimonochromatic waves through plasma. And then
the attention has been focused by many authors [14--15].

To the author's knowledge, the problem of time fractional KdV equation in
collisionless plasma has not been addressed in the literature before. So,
our motive here is to study the effects of time fractional parameter on the
electrostatic structures for a system of unmagnetized collisionless plasma
consisting of a cold electron fluid and isothermal ions with two different
temperatures obeying Boltzmann type distributions. We expect that the
inclusion of time fractional parameter will change the properties as well as
the regime of existence of solitons. Several methods have been used to solve
fractional differential equations such as: the Laplace transform method, the
Fourier transform method, the iteration method and the operational method
[16]. Recently, there are some papers deal with the existence and
multiplicity of solution of nonlinear fractional differential equation by
the use of techniques of nonlinear analysis [17-18]. In this paper, the
resultant fractional KdV equation will be solved using a
variational-iteration method (VIM) firstly used by He [19].

This paper is organized as follows: Section 2 is devoted to describe the
formulation of the time-fractional KdV (FKdV) equation using the variational
Euler-Lagrange method. In section 3, variational-Iteration Method 9VIM) is
discussed. The resultant time-FKdV equation is solved approximately using
VIM. Section 5 contains the results of calculations and discussion of these
results.

\section{Basic equations}

We consider a homogeneous system of unmagnetized collisionless plasma
consisted of a cold electron fluid and isothermal ions with two different
temperatures obeying Boltzmann type distributions. Such system is governed
by the following normalized \ equations in one dimension [15]:\quad {\large %
\quad \quad \quad \quad \quad \quad \quad \quad \quad \quad \quad \quad
\quad \quad \quad \quad \quad \quad \quad \quad \quad \quad \quad \quad
\quad \quad \quad \quad \quad \quad \quad \quad \quad \quad \quad \quad
\quad \quad \quad \quad \quad \quad \quad \quad \quad \quad \quad \quad
\quad \quad \quad \quad \quad \quad \thinspace\ \ \ \ \ \ \ \ \ \ \ \ \ \ \
\ \ \ \ \ \ \ \ \ \ \ \ \ \ \ \ \ \ \ \ \ \ \ \ \ \ \ \ \ \ \ \ \ \ \ \ \ \
\ \ \ \ \ \ \ \ \ \ \ \ \ \ \ \ \ \ \ \ \ \ \ \ \ \ \ \ \ \ \ \ \ \ \ \ \ \
\ \ \ \ \ \ \ \ \ \ \ }

\begin{eqnarray}
\frac{\partial }{\partial t}n_{e}(x,t)+\frac{\partial }{\partial x}%
[n_{e}(x,t)u_{e}(x,t)] &=&0\text{,}  \TCItag{1a} \\
\lbrack \frac{\partial }{\partial t}+u_{e}(x,t)\frac{\partial }{\partial x}%
]u_{e}(x,t)-\frac{\partial }{\partial x}\phi (x,t) &=&0\text{,}  \TCItag{1b}
\\
\frac{\partial ^{2}}{\partial x^{2}}\phi
(x,t)-n_{e}(x,t)+n_{il}(x,t)+n_{ih}(x,t) &=&0\text{,}  \TCItag{1c}
\end{eqnarray}%
the two ions density $n_{il}(x,t)$ and $n_{ih}(x,t)$\ \ are given by: 
\begin{eqnarray}
n_{il}(x,t) &=&\mu \exp [\frac{-\ \phi (x,t)}{\mu +\nu \beta }]\text{,} 
\TCItag{1d} \\
\;n_{ih}(x,t) &=&\nu \exp [\frac{-\ \beta \ \phi (x,t)}{\mu +\nu \beta }]%
\text{.}  \TCItag{1e}
\end{eqnarray}

In the earlier equations, $n_{e}(x,t)$ is the cold electron density
normalized by equilibrium value $n_{e0\text{ }}$, $u_{e}(x,t)$\ is the cold
electron fluid velocity normalized by $C_{eff}=(T_{eff}/m_{e})^{\frac{1}{2}}$%
, $T_{eff}=\frac{T_{l}T_{h}}{\mu T_{h}+\gamma T_{l}},$ $T_{l}$ is the
temperature of low temperature ion with initial normalized equilibrium
density $\mu $, $T_{h}$ is the temperature of high temperature ion with
initial normalized equilibrium density $\nu $, $\beta =\frac{T_{l}}{T_{h}}$
is the ions temperatures ratio, $\phi (x,t)$\ is the electric potential
normalized by $T_{eff}/e$, $m_{e}$\ is the mass of electron, $e$\ is the
electron charge. $x$\ is the space co-ordinate normalized to the effective
Debye length $\lambda _{D_{eff}}\ =(\frac{T_{eff}}{4\pi e^{2}n_{c0\text{ }}}%
)^{\frac{1}{2}}$ and $t$\ is the time variable normalized by the inverse of
the cold electron plasma frequency\ $\omega _{_{pc}}^{-1}$\ ,\ [$\omega
_{_{pc}}=(\frac{4\pi e^{2}n_{c0\text{ }}}{m_{e}})^{\frac{1}{2}}$]. The
neutrality condition reads $\mu +\nu =1$. Equations (1a) and (1b) represent
the inertia of cold electron and equation (1c) is the Poisson's equation
needs to make the self consistent. The two ion-densities are described by
Boltzmann type distributions given by equations (1d) and (1e).

\section{Nonlinear small-amplitude}

According to the general method of reductive perturbation theory [13], the
slow stretched co-ordinates are introduced as:

\begin{equation}
\tau =\epsilon ^{\frac{3}{2}}t,\text{ \ \ \ \ \ }\xi =\epsilon ^{\frac{1}{2}%
}(x-\lambda t),  \tag{2}
\end{equation}%
where $\epsilon $\ is a small dimensionless expansion parameter and $\lambda 
$\ is the wave speed normalized by $C_{eff}$. All physical quantities
appearing in (1) are expanded as power series in $\epsilon $\ about their
equilibrium values as:

\begin{eqnarray}
n_{e}(\xi ,\tau ) &=&1+\epsilon n_{1}(\xi ,\tau )+\epsilon ^{2}n_{2}(\xi
,\tau )+\epsilon ^{3}n_{3}(\xi ,\tau )+...\text{,}  \TCItag{3a} \\
u_{e}(\xi ,\tau ) &=&\epsilon u_{1}(\xi ,\tau )+\epsilon ^{2}u_{2}(\xi ,\tau
)+\epsilon ^{3}u_{3}(\xi ,\tau )+...\text{,}  \TCItag{3b} \\
\phi (\xi ,\tau ) &=&\epsilon \phi _{1}(\xi ,\tau )+\epsilon ^{2}\phi
_{2}(\xi ,\tau )+\epsilon ^{3}\phi _{3}(\xi ,\tau )+...\text{,}  \TCItag{3c}
\end{eqnarray}%
with the boundary conditions that as $\left\vert \xi \right\vert \rightarrow
\infty $,\ $n_{e}=1$, $u_{e}=0$, $\phi =0$.

Substituting (2) and (3) into (1) and equating the coefficients of like
powers of $\epsilon $ lead, from the lowest and second-order equations in $%
\epsilon $, to the following KdV equation for the first-order perturbed
potential:

\begin{equation}
\frac{\partial \phi _{1}(\xi ,\tau )}{\partial \tau }+A\;\phi _{1}(\xi ,\tau
)\frac{\partial \phi _{1}(\xi ,\tau )}{\partial \xi }+B\ \frac{\partial
^{3}\phi _{1}(\xi ,\tau )}{\partial \xi ^{3}}=0\text{,}  \tag{4a}
\end{equation}%
where

\begin{equation}
A=\frac{{\lambda }^{3}}{2\,}[\frac{\mu +\nu \beta \ ^{2}}{(\mu +\nu \beta )\
^{2}}-\frac{3}{{\lambda }^{4}\,}]\text{, \ }B=\frac{{\lambda }^{3}}{2\,}%
\text{ and }{\lambda =\pm 1}\text{.}  \tag{4b}
\end{equation}

In equation (4a), $\phi _{1}(\xi ,~\tau )$ is a field variable, $\xi $ is a
space coordinate in the propagation direction of the field and $\tau \in T$($%
=[0,T_{0}]$) is the time coordinate. The resultant KdV equation (4a) can be
converted into time-fractional KdV equation as follows:

Using a potential function $V(\xi ,~\tau )$, where $\phi _{1}(\xi ,~\tau
)=V_{\xi }(\xi ,~\tau )=\Phi (\xi ,\tau )$, gives the potential equation of
the regular KdV equation (4a) in the form

\begin{equation}
V_{\xi \tau }(\xi ,~\tau )+A~V_{\xi }(\xi ,~\tau )v_{\xi \xi }(\xi ,~\tau
)+B~V_{\xi \xi \xi \xi }(\xi ,~\tau )=0\text{,}  \tag{5}
\end{equation}%
where the subscripts denote the partial differentiation of the function with
respect to the parameter. The Lagrangian of this regular KdV equation (4a)
can be defined using the semi-inverse method [20, 21] as follows:

The functional of the potential equation (5) can be represented by

\begin{equation}
J(V)=\dint\limits_{R}d\xi \dint\limits_{T}d\tau \{V(\xi ,\tau )[c_{1}V_{\xi
\tau }(\xi ,\tau )+c_{2}AV_{\xi }(\xi ,\tau )v_{\xi \xi }(\xi ,\tau
)+c_{3}BV_{\xi \xi \xi \xi }(\xi ,\tau )]\}\text{,}  \tag{6}
\end{equation}%
where $c_{1}$, $c_{2}$ and $c_{3}$ are constants to be determined.
Integrating by parts and taking $V_{\tau }|_{R}=V_{\xi }|_{R}=V_{\xi
}|_{T}=0 $ lead to

\begin{equation}
J(V)=\dint\limits_{R}d\xi \dint\limits_{T}d\tau \{V(\xi ,\tau )[-c_{1}V_{\xi
}(\xi ,\tau )V_{\tau }(\xi ,\tau )-\frac{1}{2}c_{2}AV_{\xi }^{3}(\xi ,\tau
)+c_{3}BV_{\xi \xi }^{2}(\xi ,\tau )]\}\text{.}  \tag{7}
\end{equation}

The unknown constants $c_{i}(i=1$, $2$, $3)$ can be determined by taking the
variation of the functional (7) to make it optimal. Taking the variation of
this functional, integrating each term by parts and making the variation
optimum give the following relation

\begin{equation}
2c_{1}V_{\xi \tau }(\xi ,\tau )+3c_{2}AV_{\xi }(\xi ,\tau )V_{\xi \xi }(\xi
,\tau )+2c_{3}BV_{\xi \xi \xi \xi }(\xi ,\tau )=0\text{.}  \tag{8}
\end{equation}

As this equation must be equal to (5), the unknown constants are given as

\begin{equation}
c_{1}=1/2\text{, }c_{2}=1/3\text{ and }c_{3}=1/2\text{.}  \tag{9}
\end{equation}

Therefore, the functional given by (7) gives the Lagrangian of the regular
KdV equation as

\begin{equation}
L(V_{\tau },~V_{\xi },V_{\xi \xi })=-\frac{1}{2}V_{\xi }(\xi ,\tau )V_{\tau
}(\xi ,\tau )-\frac{1}{6}AV_{\xi }^{3}(\xi ,\tau )+\frac{1}{2}BV_{\xi \xi
}^{2}(\xi ,\tau )\text{.}  \tag{10}
\end{equation}

Similar to this form, the Lagrangian of the time-fractional version of the
KdV equation can be written in the form

\begin{eqnarray}
F(_{0}D_{\tau }^{\alpha }V,~V_{\xi },V_{\xi \xi }) &=&-\frac{1}{2}%
[_{0}D_{\tau }^{\alpha }V(\xi ,\tau )]V_{\xi }(\xi ,\tau )-\frac{1}{6}%
AV_{\xi }^{3}(\xi ,\tau )+\frac{1}{2}BV_{\xi \xi }^{2}(\xi ,\tau )\text{, } 
\notag \\
0 &\leq &\alpha <1\text{,}  \TCItag{11}
\end{eqnarray}%
where the fractional derivative is represented, using the left
Riemann-Liouville fractional derivative definition as [16]

\begin{eqnarray}
_{a}D_{t}^{\alpha }f(t) &=&\frac{1}{\Gamma (k-\alpha )}\frac{d^{k}}{dt^{k}}%
[\int_{a}^{t}d\tau (t-\tau )^{k-\alpha -1}f(\tau )]\text{, }  \notag \\
k-1 &\leq &\alpha \leq 1\text{, }t\in \lbrack a,b]\text{.}  \TCItag{12}
\end{eqnarray}

The functional of the time-FKdV equation can be represented in the form

\begin{equation}
J(V)=\dint\limits_{R}d\xi \dint\limits_{T}d\tau F(_{0}D_{\tau }^{\alpha
}V,~V_{\xi },V_{\xi \xi })\text{,}  \tag{13}
\end{equation}%
where the time-fractional Lagrangian $F(_{0}D_{\tau }^{\alpha }V,~V_{\xi
},V_{\xi \xi })$ is defined by (11).

Following Agrawal's method [3, 4], the variation of functional (13) with
respect to $V(\xi ,\tau )$ leads to

\begin{equation}
\delta J(V)=\dint\limits_{R}d\xi \dint\limits_{T}d\tau \{\frac{\partial F}{%
\partial _{0}D_{\tau }^{\alpha }V}\delta _{0}D_{\tau }^{\alpha }V+\frac{%
\partial F}{\partial V_{\xi }}\delta V_{\xi }+\frac{\partial F}{\partial
V_{\xi \xi }}\delta V_{\xi \xi }\}\text{.}  \tag{14}
\end{equation}

The formula for fractional integration by parts reads [3, 16]

\begin{equation}
\int_{a}^{b}dtf(t)_{a}D_{t}^{\alpha
}g(t)=\int_{a}^{t}dtg(t)_{t}D_{b}^{\alpha }f(t)\text{, \ \ \ }f(t)\text{, }%
g(t)\text{ }\in \lbrack a,~b]\text{.}  \tag{15}
\end{equation}%
where $_{t}D_{b}^{\alpha }$, the right Riemann-Liouville fractional
derivative, is defined by [16]

\begin{eqnarray}
_{t}D_{b}^{\alpha }f(t) &=&\frac{(-1)^{k}}{\Gamma (k-\alpha )}\frac{d^{k}}{%
dt^{k}}[\int_{t}^{b}d\tau (\tau -t)^{k-\alpha -1}f(\tau )]\text{, }  \notag
\\
k-1 &\leq &\alpha \leq 1\text{, }t\in \lbrack a,b]\text{.}  \TCItag{16}
\end{eqnarray}

Integrating the right-hand side of (14) by parts using formula (15) leads to

\begin{equation}
\delta J(V)=\dint\limits_{R}d\xi \dint\limits_{T}d\tau \lbrack _{\tau
}D_{T_{0}}^{\alpha }(\frac{\partial F}{\partial _{0}D_{\tau }^{\alpha }V})-%
\frac{\partial }{\partial \xi }(\frac{\partial F}{\partial V_{\xi }})+\frac{%
\partial ^{2}}{\partial \xi ^{2}}(\frac{\partial F}{\partial V_{\xi \xi }}%
)]\delta V\text{,}  \tag{17}
\end{equation}%
where it is assumed that $\delta V|_{T}=\delta V|_{R}=\delta V_{\xi }|_{R}=0$%
.

Optimizing this variation of the functional $J(V)$, i. e; $\delta J(V)=0$,
gives the Euler-Lagrange equation for the time-FKdV equation in the form

\begin{equation}
_{\tau }D_{T_{0}}^{\alpha }(\frac{\partial F}{\partial _{0}D_{\tau }^{\alpha
}V})-\frac{\partial }{\partial \xi }(\frac{\partial F}{\partial V_{\xi }})+%
\frac{\partial ^{2}}{\partial \xi ^{2}}(\frac{\partial F}{\partial V_{\xi
\xi }})=0\text{.}  \tag{18}
\end{equation}

Substituting the Lagrangian of the time-FKdV equation (11) into this
Euler-Lagrange formula (18) gives

\begin{equation}
-\frac{1}{2}~_{\tau }D_{T_{0}}^{\alpha }V_{\xi }(\xi ,\tau )+\frac{1}{2}%
~_{0}D_{\tau }^{\alpha }V_{\xi }(\xi ,\tau )+AV_{\xi }(\xi ,\tau )V_{\xi \xi
}(\xi ,\tau )+BV_{\xi \xi \xi \xi }(\xi ,\tau )=0\text{.}  \tag{19}
\end{equation}

Substituting for the potential function, $V_{\xi }(\xi ,\tau )=\phi _{1}(\xi
,\tau )=\Phi (\xi ,\tau )$, gives the time-FKdV equation for the state
function $\Phi (\xi ,\tau )$ in the form

\begin{equation}
\frac{1}{2}[_{0}D_{\tau }^{\alpha }\Phi (\xi ,\tau )-_{\tau
}D_{T_{0}}^{\alpha }\Phi (\xi ,\tau )]+A\text{ }\Phi (\xi ,\tau )\text{ }%
\Phi _{\xi }(\xi ,\tau )+B\text{ }\Phi _{\xi \xi \xi }(\xi ,\tau )=0\text{,}
\tag{20}
\end{equation}%
where the fractional derivatives $_{0}D_{\tau }^{\alpha }$ and $_{\tau
}D_{T_{0}}^{\alpha }$ are, respectively the left and right Riemann-Liouville
fractional derivatives and are defined by (12) and (16).

The time-FKdV equation represented in (20) can be rewritten by the formula

\begin{equation}
\frac{1}{2}~_{0}^{R}D_{\tau }^{\alpha }\Phi (\xi ,\tau )+A~\Phi (\xi ,\tau )%
\frac{\partial }{\partial \xi }\Phi (\xi ,\tau )+B~\frac{\partial ^{3}}{%
\partial \xi ^{3}}\Phi (\xi ,\tau )=0\text{,}  \tag{21}
\end{equation}%
where the fractional operator $_{0}^{R}D_{\tau }^{\alpha }$ is called Riesz
fractional derivative and can be represented by [4, 16]

\begin{eqnarray}
~_{0}^{R}D_{t}^{\alpha }f(t) &=&\frac{1}{2}[_{0}D_{t}^{\alpha
}f(t)+~(-1)^{k}{}_{t}D_{T_{0}}^{\alpha }f(t)]  \notag \\
&=&\frac{1}{2}\frac{1}{\Gamma (k-\alpha )}\frac{d^{k}}{dt^{k}}%
[\int_{a}^{t}d\tau |t-\tau |^{k-\alpha -1}f(\tau )]\text{, }  \notag \\
k-1 &\leq &\alpha \leq 1\text{, }t\in \lbrack a,b]\text{.}  \TCItag{22}
\end{eqnarray}

The nonlinear fractional differential equations have been solved using
different techniques [16-20]. In this paper, a variational-iteration method
(VIM) [21] has been used to solve the time-FKdV equation that is formulated
using Euler-Lagrange variational technique.

\section{Variational-iteration method}

A general Lagrange multiplier method is constructed to solve non-linear
problems, which was first proposed to solve problems in quantum mechanics
[21]. The VIM is a modification of this Lagrange multiplier method [22]. The
basic features of the VIM are as follows. The solution of the linear term of
the problem or the initial (boundary) condition of the nonlinear problem is
used as initial approximation or trail function. A more highly precise
approximation can be obtained using iteration correction functional.
Variational-iteration method (VIM) [21] has been used successfully to solve
different types of integer nonlinear differential equations [22, 23]. Also,
VIM is used to solve linear and nonlinear fractional differential equations
[24, 25]. This VIM has been used in this paper to solve the formulated
time-FKdV equation.

Considering a nonlinear partial differential equation consists of a linear
part $\overset{\symbol{94}}{L}U(x,t)$, nonlinear part $\overset{\symbol{94}}{%
N}U(x,t)$ and a free term $f(x,t)$ represented as

\begin{equation}
\overset{\symbol{94}}{L}U(x,t)+\overset{\symbol{94}}{N}U(x,t)=f(x,t)\text{,}
\tag{23}
\end{equation}%
where $\overset{\symbol{94}}{L}$ is the linear operator and $\overset{%
\symbol{94}}{N}$ is the nonlinear operator. According to the VIM, the ($n+1$)%
\underline{th} approximation solution of (23) can be given by the iteration
correction functional as [24, 25]

\begin{equation}
U_{n+1}(x,t)=U_{n}(x,t)+\int_{0}^{t}d\tau \lambda (\tau )[\overset{\symbol{94%
}}{L}U_{n}(x,\tau )+\overset{\symbol{94}}{N}\hat{U}_{n}(x,\tau )-f(x,\tau )]%
\text{, }n\geq 0\text{,}  \tag{24}
\end{equation}%
where $\lambda (\tau )$ is a Lagrangian multiplier and $\hat{U}_{n}(x,\tau )$
is considered as a restricted variation function, i. e; $\delta \hat{U}%
_{n}(x,\tau )=0$. Extreme the variation of the correction functional (24)
leads to the Lagrangian multiplier $\lambda (\tau )$. The initial iteration
can be used as the solution of the linear part of (23) or the initial value $%
U(x,0)$. As $n$ tends to infinity, the iteration leads to the exact solution
of (23), i. e;

\begin{equation}
U(x,t)=\underset{n\rightarrow \infty }{\lim }U_{n}(x,t)\text{.}  \tag{25}
\end{equation}%
\qquad For linear problems, the exact solution can be given using this
method in only one step where its Lagrangian multiplier can be exactly
identified.

\section{Time-fractional KdV equation solution}

The time-FKdV equation represented by (21) can be solved using the VIM by
the iteration correction functional (24) as follows:

Affecting from left by the fractional operator $_{0}^{R}D_{\tau }^{\ \alpha
-1}$ on (21) leads to

\begin{eqnarray}
\frac{\partial }{\partial \tau }\Phi (\xi ,\tau ) &=&~_{0}^{R}D_{\tau }^{\
\alpha -1}\Phi (\xi ,\tau )|_{\tau =0}\frac{\tau ^{\alpha -2}}{\Gamma
(\alpha -1)}  \notag \\
&&-\ _{0}^{R}D_{\tau }^{\ 1-\alpha }[A~\Phi (\xi ,\tau )\frac{\partial }{%
\partial \xi }\Phi (\xi ,\tau )+B~\frac{\partial ^{3}}{\partial \xi ^{3}}%
\Phi (\xi ,\tau )]\text{, }  \notag \\
0 &\leq &\alpha \leq 1\text{, }\tau \in \lbrack 0,T_{0}]\text{,}  \TCItag{26}
\end{eqnarray}%
where the following fractional derivative property is used [16]

\begin{equation}
\ _{a}^{R}D_{b}^{\ \alpha }[\ _{a}^{R}D_{b}^{\ \beta
}f(t)]=~_{a}^{R}D_{b}^{\ \alpha +\beta }f(t)-\overset{k}{\underset{j=1}{\sum 
}}\ _{a}^{R}D_{b}^{\ \beta -j}f(t)|_{t=a}~\frac{(t-a)^{-\alpha -j}}{\Gamma
(1-\alpha -j)}\text{, }k-1\leq \beta <k\text{.}  \tag{27}
\end{equation}

As $\alpha <1$, the Riesz fractional derivative $_{0}^{R}D_{\tau }^{\ \alpha
-1}$ is considered as Riesz fractional integral $_{0}^{R}I_{\tau }^{1-\alpha
}$ that is defined by [16]

\begin{equation}
\ _{0}^{R}I_{\tau }^{\ \alpha }f(t)=\frac{1}{2}[_{0}I_{\tau }^{\ \alpha
}f(t)\ +~_{\tau }I_{b}^{\ \alpha }f(t)]=\frac{1}{2}\frac{1}{\Gamma (\alpha )}%
\int_{a}^{b}d\tau |t-\tau |^{\alpha -1}f(\tau )\text{, }\alpha >0\text{,} 
\tag{28}
\end{equation}%
where $_{0}I_{\tau }^{\ \alpha }f(t)$\ and $_{\tau }I_{b}^{\ \alpha }f(t)$
are the left and right Riemann-Liouvulle fractional integrals, respectively
[16].

The iterative correction functional of equation (26) is given as

\begin{eqnarray}
\Phi _{n+1}(\xi ,\tau ) &=&\Phi _{n}(\xi ,\tau )+\int_{0}^{\tau }d\tau
^{\prime }\lambda (\tau ^{\prime })\{\frac{\partial }{\partial \tau ^{\prime
}}\Phi _{n}(\xi ,\tau ^{\prime })  \notag \\
&&-~_{0}^{R}I_{\tau ^{\prime }}^{1-\alpha }\Phi _{n}(\xi ,\tau ^{\prime
})|_{\tau ^{\prime }=0}\frac{\tau ^{\prime \alpha -2}}{\Gamma (\alpha -1)} 
\notag \\
&&+\ _{0}^{R}D_{\tau ^{\prime }}^{\ 1-\alpha }[A~\overset{\symbol{126}}{\Phi
_{n}}(\xi ,\tau ^{\prime })\frac{\partial }{\partial \xi }\overset{\symbol{%
126}}{\Phi _{n}}(\xi ,\tau ^{\prime })+B~\frac{\partial ^{3}}{\partial \xi
^{3}}\overset{\symbol{126}}{\Phi _{n}}(\xi ,\tau ^{\prime })]\}\text{,} 
\TCItag{29}
\end{eqnarray}%
where $n\geq 0$ and the function $\overset{\symbol{126}}{\Phi _{n}}(\xi
,\tau )$ is considered as a restricted variation function, i. e; $\delta 
\overset{\symbol{126}}{\Phi _{n}}(\xi ,\tau )=0$. The extreme of the
variation of (29) using the restricted variation function leads to

\begin{eqnarray*}
\delta \Phi _{n+1}(\xi ,\tau ) &=&\delta \Phi _{n}(\xi ,\tau
)+\int_{0}^{\tau }d\tau ^{\prime }\lambda (\tau ^{\prime })~\delta \frac{%
\partial }{\partial \tau ^{\prime }}\Phi _{n}(\xi ,\tau ^{\prime }) \\
&=&\delta \Phi _{n}(\xi ,\tau )+\lambda (\tau )~\delta \Phi _{n}(\xi ,\tau
)-\int_{0}^{\tau }d\tau ^{\prime }\frac{\partial }{\partial \tau ^{\prime }}%
\lambda (\tau ^{\prime })~\delta \Phi _{n}(\xi ,\tau ^{\prime })=0\text{.}
\end{eqnarray*}

This relation leads to the stationary conditions $1+\lambda (\tau )=0$ and $%
\frac{\partial }{\partial \tau ^{\prime }}\lambda (\tau ^{\prime })=0$,
which leads to the Lagrangian multiplier as $\lambda (\tau ^{\prime })=-1$.
\ Therefore, the correction functional (29) is given by the form

\begin{eqnarray}
\Phi _{n+1}(\xi ,\tau ) &=&\Phi _{n}(\xi ,\tau )-\int_{0}^{\tau }d\tau
^{\prime }\{\frac{\partial }{\partial \tau ^{\prime }}\Phi _{n}(\xi ,\tau
^{\prime })  \notag \\
&&-~_{0}^{R}I_{\tau ^{\prime }}^{1-\alpha }\Phi _{n}(\xi ,\tau ^{\prime
})|_{\tau ^{\prime }=0}\frac{\tau ^{\prime \alpha -2}}{\Gamma (\alpha -1)} 
\notag \\
&&+\ _{0}^{R}D_{\tau ^{\prime }}^{\ 1-\alpha }[A~\Phi _{n}(\xi ,\tau
^{\prime })\frac{\partial }{\partial \xi }\Phi _{n}(\xi ,\tau ^{\prime })+B~%
\frac{\partial ^{3}}{\partial \xi ^{3}}\Phi _{n}(\xi ,\tau ^{\prime })]\}%
\text{,}  \TCItag{30}
\end{eqnarray}%
where $n\geq 0$.

In Physics, if $\tau $ denotes the time-variable, the right
Riemann-Liouville fractional derivative is interpreted as a future state of
the process. For this reason, the right-derivative is usually neglected in
applications, when the present state of the process does not depend on the
results of the future development [3]. Therefore, the right-derivative is
used equal to zero in the following calculations.

The zero order correction of the solution can be taken as the initial value
of the state variable, which is taken in this case as

\begin{equation}
\Phi _{0}(\xi ,\tau )=\Phi (\xi ,0)=A_{0}\sec \text{h}^{2}(c\xi )\text{.} 
\tag{31}
\end{equation}

where $A_{0}=\frac{3v}{A}$ and $c=\frac{1}{2}\sqrt{\frac{v}{B}}$ are
constants.

Substituting this zero order approximation into (30) and using the
definition of the fractional derivative (22) lead to the first order
approximation as

\begin{eqnarray}
\Phi _{1}(\xi ,\tau ) &=&A_{0}\sec \text{h}(c\xi )^{2}+2A_{0}c~\sinh (c\xi
)~\sec \text{h}(c\xi )^{3}  \notag \\
&&\ast \lbrack 4c^{2}B+(A_{0}A-12c^{2}B)\sec \text{h}(c\xi )^{2}]\frac{\tau
^{\alpha }}{\Gamma (\alpha +1)}\text{.}  \TCItag{32}
\end{eqnarray}

Substituting this equation into (30), using the definition (22) and the
Maple package lead to the second order approximation in the form

\begin{eqnarray}
\Phi _{2}(\xi ,\tau ) &=&A_{0}\sec \text{h}(c\xi )^{2}+2A_{0}c~\sinh (c\xi
)~\sec \text{h}(c\xi )^{3}  \notag \\
&&\ast \lbrack 4c^{2}B+(A_{0}A-12c^{2}B)\sec \text{h}(c\xi )^{2}]\frac{\tau
^{\alpha }}{\Gamma (\alpha +1)}  \notag \\
&&+2A_{0}c^{2}\sec \text{h}(c\xi )^{2}  \notag \\
&&\ast \lbrack 32c^{4}B^{2}+16c^{2}B(5A_{0}A-63c^{2}B)\sec \text{h}(c\xi
)^{2}  \notag \\
&&+2(3A_{0}^{2}A^{2}-176A_{0}c^{2}AB+1680c^{4}B^{2})\sec \text{h}(c\xi )^{4}
\notag \\
&&-7(A_{0}^{2}A^{2}-42A_{0}c^{2}AB+360c^{4}B^{2})\sec \text{h}(c\xi )^{6}]%
\frac{\tau ^{2\alpha }}{\Gamma (2\alpha +1)}  \notag \\
&&+4A_{0}^{2}c^{3}\sinh (c\xi )~\sec \text{h}(c\xi )^{5}  \notag \\
&&\ast \lbrack 32c^{4}B^{2}+24c^{2}B(A_{0}A-14c^{2}B)\sec \text{h}(c\xi )^{2}
\notag \\
&&+4(A_{0}^{2}A^{2}-32A_{0}c^{2}AB+240c^{4}B^{2})\sec \text{h}(c\xi )^{4} 
\notag \\
&&-5(A_{0}^{2}A^{2}-24A_{0}c^{2}AB+144c^{4}B^{2})\sec \text{h}(c\xi )^{6}] 
\notag \\
&&\ast \frac{\Gamma (2\alpha +1)}{[\Gamma (\alpha +1)]^{2}}\frac{\tau
^{3\alpha }}{\Gamma (3\alpha +1)}\text{.}  \TCItag{33}
\end{eqnarray}

The higher order approximations can be calculated using the Maple or the
Mathematica package to the appropriate order where the infinite
approximation leads to the exact solution.

\section{Results and discussion}

Numerical studies have been made for a small amplitude electron-acoustic
waves in an unmagnetized collisionless plasma consisted of a cold electron
fluid and isothermal ions with two different temperatures obeying Boltzmann
type distributions. We have derived the Korteweg-de Vries equation by using
the reductive perturbation method [13]. The Riemann-Liouvulle fractional
derivative [16] is used to describe the time fractional operator in the FKdV
equation. He's variational-iteration method [21] is used to solve the
derived time-FKdV equation.

However, since one of our motivations was to study effects of initial
equilibrium density $\mu $ of low temperature ion and time fractional order $%
\alpha $ on the existence of solitary waves. Our system can support two
kinds of potential structure namely, compressive and rarefactive pulses.
Depending on the sign of the coefficient of the nonlinear term $A$,
compressive soliton exists if $A>0$ while rarefactive soliton exists if $A<0$%
.

In Fig (1), profiles of the bell-shaped rarefactive and compressive solitary
pulses are obtained due to the change of the range of . Figure (2) shows
that both the amplitude and the width of the compressive solitons increase
with while both decrease for rarefactive solitons. Also, the time fractional
order decreases the amplitude of the rarefactive and compressive solitons as
shown in Fig (3).

In summery, it has been found that amplitude and width of the
electron-acoustic waves as well as parametric regime where the solitons can
exist is sensitive to the initial equilibrium density of low temperature ion
. Moreover, the time fractional order plays the role of higher order
perturbation theory in changing the soliton amplitude.

The application of our model might be particularly interesting in the new
observations for the Earth's plasma sheet boundary layer region. We have
stressed out that it is necessary to study the critical case for $A=0,$%
{\large \ }the amplitude of the solitary pulse tends to infinity and the
time-FKdV\ equation is not appropriate for describing the system. This is
beyond the scope of the further work.\pagebreak

\textbf{Figure Captions}

\textbf{Fig (1):} The electric potential $\Phi (\xi ,\tau )$ vs $\xi $ and $%
\tau $ for $\lambda =1$, $v=0.04$, $\alpha =0.5$, $\beta =0.05$, (a) $\mu
=0.2$ and (b) $\mu =0.3$.

\textbf{Fig (2):} The electric potential $\Phi (\xi ,\tau )$ vs $\xi $ and $%
\mu $ for $\lambda =1$, $v=0.04$, $\alpha =0.5$, $\beta =0.05$ and $\tau =5$%
: (a) 3 dimensions and (b) 2 dimensions.

\textbf{Fig (3):} The amplitude of the electric potential $\Phi (0,\tau )$
vs $\tau $ and $\alpha $ for $\lambda =1$, $v=0.04$, $\beta =0.05$ and $\mu
=0.2$: (a) 3 dimensions and (b) 2 dimensions.\pagebreak

\end{document}